\documentclass[12pt]{article}
\def\beqa{\begin{eqnarray}}
\def\eeqa{\end{eqnarray}}

\usepackage{graphics}
\begin{document}
\begin{titlepage}
         \title{ Maximal Acceleration Effects in Kerr
         Space.}
\author{V. Bozza$^{a,b}$\thanks{E-mail: bozza,lambiase,scarpetta@sa.infn.it},
A. Feoli$^{b,c}$\thanks{E-mail: feoli@unisannio.it}, G.
Lambiase$^{a, b}$,
 G.Papini$^{d, f}$\thanks{E-mail: papini@uregina.ca}, G.Scarpetta$^{b,e,f}$ \\
 {\em $^a$Dipartimento di Scienze Fisiche ''E.R.Caianiello'' Universit\'a di Salerno} \\
   {\em  84081 - Baronissi - Salerno} \\
 {\em $^b$Istituto Nazionale di Fisica Nucleare, Sez. di
 Napoli, Italy.} \\
 {\em $^c$Facolt\'a d'Ingegneria, Universit\'a del Sannio} \\
 {\em $^d$Department of Physics, University of Regina,} \\
 {\em Regina, Sask. S4S 0A2, Canada.} \\
 {\em $^e$Dipartimento di Fisica, Universit\'a di Salerno, 84081 Baronissi (Sa),
        Italy.} \\
 {\em $^f$International Institute for Advanced Scientific
 Studies,} \\
 {\em  Vietri sul Mare (Sa), Italy.}
 }
              \date{\today}
              \maketitle

              \begin{abstract}
We consider a model in which accelerated particles experience line--elements with maximal
acceleration corrections that are introduced by means of successive approximations. It is
shown that approximations higher than the first need not be considered. The method is
then applied to the Kerr metric. The effective field experienced by accelerated test
particles contains corrections that vanish in the limit $\hbar\to 0$, but otherwise
affect the behaviour of matter greatly. The corrections generate potential barriers that
are external to the horizon and are impervious to classical particles.
 \end{abstract}
\thispagestyle{empty} \vspace{20. mm}

              \vfill
          \end{titlepage}

This work is concerned with a geometrical model of quantum mechanics proposed by
Caianiello \cite{qg}. The model interprets quantization as curvature of the relativistic
eight--dimensional space--time tangent bundle $TM = M_{4}\otimes TM_{4}$ ($M_4$ is the
usual flat space--time manifold of metric $\eta_{\mu\nu}$), satisfies the Born
reciprocity principle and incorporates the notion that the proper accelerations of
massive particles along their worldlines are normalized to an upper limit ${\cal A}_m$
\cite{ma}, referred to as maximal acceleration (MA). The value of  ${\cal A}_m$ can be
derived from quantum mechanical considerations \cite{ca}, \cite{pw}. Classical and
quantum arguments supporting the existence of a MA have been frequently advanced
\cite{prove},\cite{wh},\cite{b}. MA also appears in the context of Weyl space \cite{pap}
and of a geometrical analogue of Vigier'stochastic theory \cite{jv} and plays a role in
numerous issues. It is invoked as a tool to rid black hole entropy of ultraviolet
divergences \cite{McG} and of inconsistencies stemming from the application of the
point-like concept to relativistic particles \cite{he}. MA may be also regarded as a
regularization procedure, alternative  to those in which space--time is quantized by
means of a fundamental length \cite{qs}. The advantage of Caianiello's proposal here lies
in the preservation of the continuum structure of space-time.

An upper limit to the acceleration also exists in string theory where Jeans--like
instabilities \cite{gsv} occur \cite{gasp} when the acceleration induced by the
background gravitational field is large enough to render the string extremities causally
disconnected. This critical acceleration $a_c$ is determined by the string size $\lambda$
and is given by $a_c = \lambda^{-1} = (m\alpha)^{-1}$ where $m$ is the string mass and
$\alpha^{-1}$ the usual string tension.

Frolov and Sanchez \cite{fs} have found that a universal critical acceleration $a_c
\simeq \lambda^{-1}$ must be a general property of strings. The acceleration cut--off is
the same required by Sanchez in order to regularize the entropy and the free energy of
quantum strings \cite{sa2}.

In all these instances the critical acceleration is a consequence of the interplay of the
Rindler horizon with the finite extension of the particle. In Caianiello's proposal the
maximal proper acceleration is a basic physical property of all massive particles, which
is an inescapable consequence of quantum mechanics \cite{ca}, \cite{pw}, and must
therefore be included in the physical laws from the outset . This requires a modification
of the metric structure of space-time. It leads, in the case of Rindler space, to a
manifold with a non vanishing scalar curvature and a shift in the horizon \cite{emb}.

Applications of Caianiello's model include cosmology \cite{infl}, where the initial
singularity can be avoided while preserving inflation, the dynamics of accelerated
strings \cite{Feo}, the energy spectrum of a uniformly accelerated particle \cite{emb}
and neutrino oscillations \cite{8},\cite{qua}. The model also makes the metric
observer--dependent, as conjectured by Gibbons and Hawking \cite{Haw}.

The extreme large value that ${\cal A}_m=2m c^3/\hbar$ takes for
all known particles makes a direct test of the model very
difficult. Nonetheless a direct test that uses photons in a cavity
has also been suggested \cite{15}.

We have worked out the consequences of the model for the
classical electrodynamics of a particle \cite{cla}, the mass of
the Higgs boson \cite{Higgs} and the Lamb shift in hydrogenic
atoms \cite{lamb}. In the last instance the agreement between
experimental data and MA corrections is very good for $H$ and
$D$. For $He^+$ the agreement between theory and experiment is
improved by $50\%$ when MA corrections are included.

MA effects in muonic atoms appear to be measurable \cite{muo}. MA
also affects the helicity and chirality of particles \cite{chen}.

More recently, we have applied the model to the falling of massive particles in the
gravitational field of a spherically symmetric collapsing object \cite{sch}. In this
problem MA manifests itself through a spherical shell external to the Schwarzschild
horizon and impenetrable to classical particles. Massive, spinless bosons do not fare
better \cite{boson}. Nor is the shell a sheer product of the coordinate system. It does
survive, for instance, in isotropic coordinates. It is also present in the
Reissner-Nordstr\"om case \cite{reiss}. The usual process of formation of a black hole
does not therefore appear viable in the model.

In this work we examine the possibility that the formation of the barrier at the horizon
be a construct of the iteration procedure adopted \cite{sch},\cite{boson}. This is the
first objective of the paper. The second objective deals with the angular momentum of the
source which has so far been neglected. In fact, a collapsing object would very likely
possess some angular momentum. One would then like to know whether some of the MA effects
found persist in the case of the Kerr metric.

The embedding procedure introduced in \cite{sch} stipulates that the line element
experienced by an accelerating particle is represented by

\begin{equation} \label{eq1}
d\tau^2=\left(1+\frac{g_{\mu\nu}\ddot{x}^{\mu}\ddot{x}^{\nu}}{{\cal A}_m^2}
\right)g_{\alpha\beta}dx^{\alpha}dx^{\beta}\equiv \sigma^2(x)
g_{\alpha\beta}dx^{\alpha}dx^{\beta}\,.
\end{equation}
As a consequence, the effective space-time geometry experienced by accelerating particles
exhibits mass-dependent corrections that in general induce curvature and give rise to
mass-dependent violations of the equivalence principle. The four--acceleration $\ddot
x^\mu = d^2 x^\mu/d\,s^2$ appearing in (\ref{eq1}) is a rigorously covariant quantity
only for linear coordinate transformations. Though its transformation properties are
known, $\ddot x^\mu$ is in general neither covariant nor necessarily orthogonal to the
four--velocity $\dot x^\mu$, as in Minkowski space. The justification for this choice
lies primarily with the quantum mechanical derivation of ${\cal A}_m$ which applies to
$\ddot x^\mu$, is Newtonian in spirit (it requires the notion of force) and is only
compatible with special relativity. No extension of this derivation to general relativity
has so far been given. The choice of $\ddot x^\mu$ in (\ref{eq1}) is, of course,
supported by the weak field approximation to $g_{\mu\nu}$
 which, to first order, is entirely Minkowskian. Estimates of $\ddot x^\mu$
 derived below assume that MA effects represent only perturbations of the normal
 particle motion represented by geodesics. These are described by fully covariant
 equations.
 In order to compare their results, any two observers would then determine each other's
$\ddot x^\mu$ and $\sigma^2$ from their relative motion and the geodesics for a particle
of the same mass in each other's frame. Lack of covariance is not therefore fatal in this
respect. Other relevant points must be made. The model introduced is not intended to
supplant general relativity, but only to provide a method to calculate the MA corrections
to a Kerr line element. The effective gravitational field introduced in (\ref{eq1}) can
not be easily incorporated in general relativity (it violates, for one, the equivalence
principle). Nor are the symmetries of general relativity indiscriminately applicable to
(\ref{eq1}). For instance, the conformal factor is not an invariant, nor can it be
eliminated by means of general coordinate transformations. The embedding procedure
requires that it be present and that it be calculated in the same coordinates of the
unperturbed gravitational background. On the other hand, Einstein's equivalence principle
does not carry through readily to the quantum level \cite{lamm}, \cite{singh} and the
same may be expected of its consequences, like the principle of general covariance
\cite{wein}. Spectacular observer-dependent quantum mechanical effects are discussed by
Gibbons and Hawking \cite{Haw}. Complete covariance is, of course, restored in the limit
$\hbar\to 0$, whereby all quantum corrections, including those due to MA, vanish. It is
essential to keep these distinctions in mind in what follows.

The acceleration of a particle in Schwarzschild space diverges in proximity of the
gravitational radius. A careful investigation of the embedding procedure is therefore
necessary in order to better understand the validity of the approximation. For
convenience, the units $\hbar =c=G=1$ are used below.

\begin{figure}
\begin{center}
\resizebox{7cm}{!}{\includegraphics{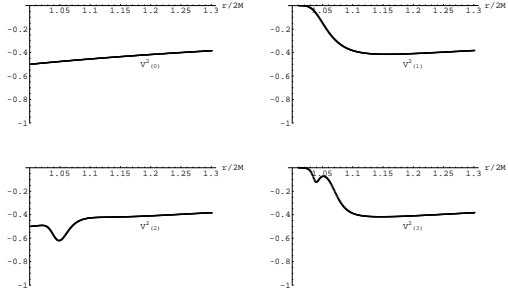}} \hfill \caption{
\footnotesize {The classical potential $V_{(0)}^2$ and the
effective potentials from the first three embeddings.}}
\label{Emb0123}
\end{center}
\end{figure}

\begin{figure}
\begin{center}
\resizebox{7cm}{!}{\includegraphics{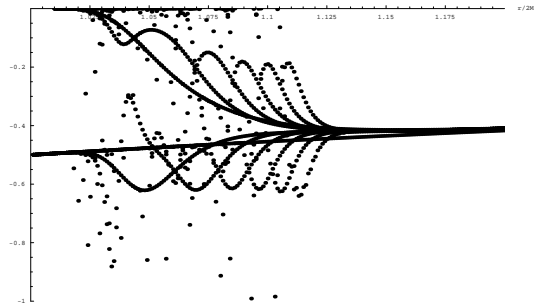}} \hfill \caption{
\footnotesize {The oscillating behaviour of $V^2_{n}$ up to
$n=12$.}} \label{Succemb}
\end{center}
\end{figure}

\begin{figure}
\begin{center}
\resizebox{7cm}{!}{\includegraphics{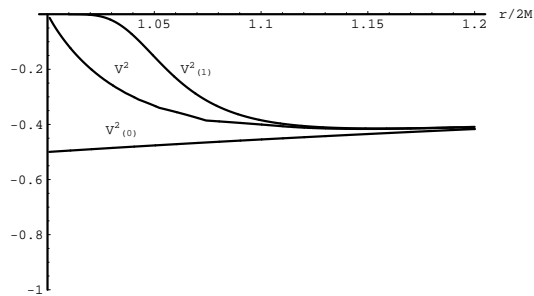}} \hfill \caption{
\footnotesize {A numerical comparison of $V^2$ with $V_{(0)}^2$
and $V_{(1)}^2$ for $E=1$ and $L=0$.}} \label{Exacext}
\end{center}
\end{figure}

\begin{figure}
\begin{center}
\resizebox{7cm}{!}{\includegraphics{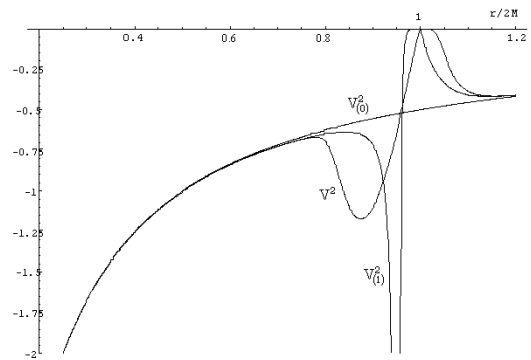}} \hfill \caption{
\footnotesize {Classical and effective potentials inside the
Schwarzschild sphere.}} \label{Exacint}
\end{center}
\end{figure}

\begin{figure}
\begin{center}
\resizebox{7cm}{!}{\includegraphics{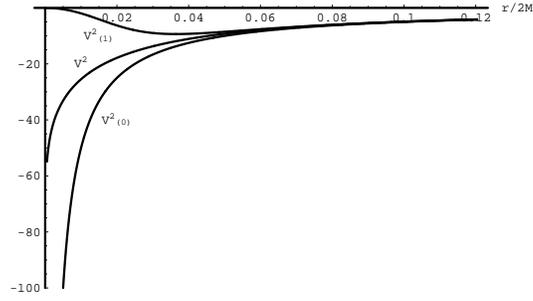}} \hfill \caption{
\footnotesize {Behaviour of $V_{(0)}^2$, $V_{(1)}^2$ and
$V_{(2)}^2$ near the origin.}} \label{Exacr=0}
\end{center}
\end{figure}

\begin{figure}
\begin{center}
\resizebox{7cm}{!}{\includegraphics{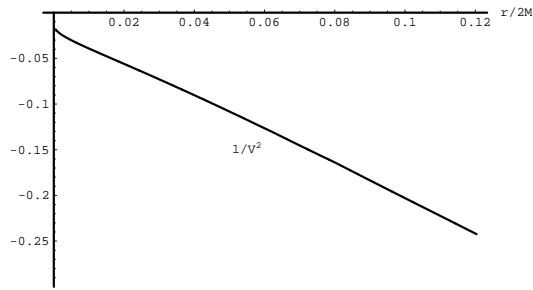}} \hfill
\caption{ \footnotesize {The plot of $V^{-2}$ near the origin
indicates that the exact effective potential $V^2$ converges for
$r\to 0$.}} \label{Invexr=0}
\end{center}
\end{figure}

The Lagrangian of a particle in a metric isotropically conformal to that of Schwarzschild
is
\begin{equation}
{\cal L}=-\frac{1}{2}\sigma ^{2}\left( r\right) \left[ \left(
1-\frac{2M}{r}\right) \dot{t}^{2}-\left( 1-\frac{2M}{r}\right)
^{-1}\dot{r}^{2}-r^{2} \dot{\varphi}^{2}\right]
\end{equation}
and the 4-acceleration is
\begin{equation}
g_{\mu \nu }\ddot{x}^{\mu }\ddot{x}^{\nu }=F\left( r,\sigma
^{2}\left( r\right) \right)\,, \label{Acceleration}
\end{equation}
where
\begin{equation} \label{F(r)}
F\left( r,\sigma ^{2}\left( r\right) \right) =\left\{ \left( 1-\frac{2M}{r}%
\right) \left[ \frac{2ME}{r^{2}\left( 1-\frac{2M}{r}\right) ^{2}\sigma ^{2}}+%
\frac{E}{\left( 1-\frac{2M}{r}\right) \sigma ^{4}}\frac{d\sigma ^{2}}{dr}%
\right] ^{2}\right.
\end{equation}
 $$
\left. -r^{2}\left( \frac{2L}{r^{3}\sigma ^{2}}+\frac{L}{r^{2}\sigma ^{4}}%
\frac{d\sigma ^{2}}{dr}\right) ^{2}\right\} \left[ \frac{E^{2}}{\sigma ^{4}}%
-\left( 1-\frac{2M}{r}\right) \left( \frac{1}{\sigma ^{2}}+\frac{L^{2}}{%
r^{2}\sigma ^{4}}\right) \right]
 $$
 $$
 -\frac{1}{\left( 1-\frac{2M}{r}\right) }\left\{ -\frac{M}{r^{2}\sigma ^{2}}+%
\frac{L^{2}}{r^{3}\sigma ^{4}}-\frac{3ML^{2}}{r^{4}\sigma
^{4}}\right. \\
\left. -\left[ \frac{E^{2}}{\sigma ^{6}}-\left(
1-\frac{2M}{r}\right) \left( \frac{1}{2\sigma
^{4}}+\frac{L^{2}}{r^{2}\sigma ^{6}}\right) \right] \frac{
d\sigma ^{2}}{dr}\right\} ^{2}\,,
 $$
$M$ is the mass of the source, $E$ the energy of the particle and $L$ its angular
momentum. Setting $\sigma^{2}=1$ in (\ref{Acceleration}), one recovers the classical
expression $g_{\mu \nu}\ddot{x}_{\left( 0\right) }^{\mu }\ddot{x}_{\left( 0\right) }^{\nu
}$ which is used in the first embedding to construct
\begin{equation}\label{omegafirst}
\sigma _{\left( 1\right) }^{2}\left( r\right) =\left(
1+\frac{g_{\mu \nu } \ddot{x}_{\left( 0\right) }^{\mu
}\ddot{x}_{\left( 0\right) }^{\nu }}{{\cal A}^2_m}\right)\,.
\end{equation}
The dynamics generated by $\sigma _{\left( 1\right) }^{2}$ yields a new quantity $g_{\mu
\nu}\ddot{x}_{\left( 1\right) }^{\mu }\ddot{x} _{\left( 1\right) }^{\nu }$, given by
(\ref{Acceleration}) with $\sigma ^{2}=\sigma _{\left( 1\right) }^{2}$, which already
contains the MA corrections and can be used to build the conformal factor of the second
embedding
\begin{equation}\label{2omega}
\sigma _{\left( 2\right) }^{2}\left( r\right) =\left(
1+\frac{g_{\mu \nu } \ddot{x}_{\left( 1\right) }^{\mu
}\ddot{x}_{\left( 1\right) }^{\nu }}{{\cal A}^{2}_m}\right)\,.
\end{equation}
Eq.(\ref{2omega}) can then be used to calculate $g_{\mu \nu }\ddot{x}_{\left( 2\right)
}^{\mu }\ddot{x}_{\left( 2\right) }^{\nu }$ by means of (\ref{Acceleration}), and so on.
In particular
\begin{equation}
\sigma _{\left( n+1\right) }^{2}=G\left( r,\sigma _{\left(
n\right) }^{2}\right) \label{(n+1) conformal factor}\,,
\end{equation}
where
\begin{equation}
G\left( r,\sigma ^{2}\right) =\left( 1+\frac{F\left( r,\sigma
^{2}\right) }{ {\cal A}^{2}_m}\right)\,.  \label{G(r)}
\end{equation}
The effects of MA can be studied through the effective potential, defined by
\begin{equation}
\left( \frac{dr}{ds}\right) ^{2}=E^{2}-V_{eff}^{2}.
\end{equation}
One finds
\begin{equation}
V_{eff}^{2}\left( r,\sigma ^{2}\left( r\right) \right)
=E^{2}-\frac{E^{2}}{ \sigma ^{4}\left( r\right) }+\left(
1-\frac{2M}{r}\right) \left( \frac{1}{ \sigma ^{2}\left( r\right)
}+\frac{L^{2}}{r^{2}\sigma ^{4}\left( r\right) } \right)
\label{Effective potentiale}\,.
\end{equation}
Eqs. (\ref{(n+1) conformal factor}), (\ref{G(r)}) and (\ref{F(r)}) give the successive
embeddings with initial condition $\sigma _{\left( 0\right) }^{2}=1$. The resulting
effects on the dynamics of the particle can be analyzed numerically by means of
(\ref{Effective potentiale}). Fig.\ref{Emb0123}a shows the classical effective potential
for a particle with $E=1 $ and $L=0$ in the region immediately external to the
gravitational radius. Figs.\ref{Emb0123}b, c, d show the effective potentials in the
first, second and third embeddings respectively. In these plots the potential barrier of
the first embedding disappears in the second, but reappears in the third. An analysis of
the function $G\left( r,\sigma^{2}\right) $ explains this behaviour. In the even
embeddings, $\sigma ^{2}$ tends in fact to unity at $r = 2M$. The presence of inverse powers of $\left( 1-\frac{%
2M}{r}\right) $ causes the odd $\sigma ^{2}$'s to diverge as $\left( 1-\frac{2M%
}{r}\right) ^{-3}$. Inversely, if $\sigma ^{2}$ diverges as $\left( 1-\frac{2M}{r}\right)
^{-3}$ in the odd embeddings, then $F\left( r,\sigma ^{2}\right) $ vanishes and $\sigma
^{2}$ is unity in the even embeddings. So the conformal factor alternates divergent to
regular behaviour. Consequently, $V_{eff}^{2}$ equals the classical potential in even
embeddings and tends to $E^{2}$ in the odd ones. This rules out any possibility that the
iteration procedure converge at $r=2M$. Fig.\ref{Succemb} indicates that instability
could even extend to outer regions in the form of an increasing, oscillating behaviour.
This instability requires a closer study. {\it It will become clear below that the first
embedding can be used as a good approximation in the model}. The starting point of the
analysis is Eq. (\ref{(n+1) conformal factor}). If a new conformal factor is calculated
by using the acceleration of a dynamics in which $\sigma^2$ is the conformal factor, one
must again find $\sigma ^{2}$. Then the correct solution is characterized as a fixed
point of (\ref{(n+1) conformal factor}). To gain information about the behaviour of
$\sigma ^{2}$ at $r = 2M$, it is useful to inspect the divergence order. This is
accomplished by writing
\begin{equation}\label{omegalfa}
\sigma ^{2}=f\left( r\right) \left( 1-\frac{2M}{r}\right) ^{\alpha
}\,,
\end{equation}
where $f\left( r\right)$ is regular at $r=2M$, and substituting (\ref{omegalfa}) into the
fixed point equation
\begin{equation}
\sigma ^{2}=G\left( r,\sigma ^{2}\right)\,.
\end{equation}
Collecting terms with the same value of $\alpha$, the two sides of
the equation are consistent with each other if and only if the
divergence orders of the
leading terms are the same. This condition is only fulfilled by $\alpha =-%
\frac{3}{5}$. If $\sigma ^{2}$ diverges, then the effective potential tends to $%
E^{2}$ and the barrier forms. In order to get a complete view of the behaviour of the
effective potential, one can resort to numerical algorithms. If one starts from any test
value $\tilde{\sigma}^{2}$, $G\left( r,\tilde{\sigma}^{2}\right) $ indicates
whether $\tilde{\sigma}^{2}$ is greater or less than the "exact" solution $%
\sigma ^{2}$. By decreasing the test value in the first case and increasing it in the
second case, one can reach the exact $\sigma ^{2}$ with an arbitrary degree of accuracy.
By building $\sigma ^{2}$ far from $r = 2M$ so that $\sigma ^{2}=1$ is a good initial
condition, one gradually approaches $r = 2M$ always replacing derivatives with
incremental ratios. In Fig.\ref {Exacext} the effective potential resulting from this
numerical approach is compared to the classical potential and the result of the first
embedding for a particle with $E=1$ and $L=0$. {\it The presence of a barrier at $r=2M$
is confirmed}, even though
the lower degree of divergence of the exact $\sigma ^{2}$ with respect to $%
\sigma _{\left( 1\right) }^{2}$ produces a different behaviour near the horizon. To
investigate the correct solution inside the Schwarzschild sphere, one must first overcome
the divergence in $\sigma ^{2}$. If one applies the same algorithm to $\Sigma ^{2}\left(
r\right) =\frac{1}{\sigma ^{2}\left( r\right) }$, the divergence at $r=2M$ is replaced by
zero and the interior of the black hole can be studied. Fig.\ref{Exacint} compares the
effective potentials for $r<2M$. The singularity in the first embedding is absent in the
exact numerical solution which instead has a simple minimum. Notice that the exact
$\sigma ^{2}$ is everywhere positive while $\sigma _{\left( 1\right) }^{2}$ is negative
between the singularity and $2M$. Finally, Fig.\ref{Exacr=0} indicates that the exact
potential does not vanish at $r = 0$. A plot of $1/V_{eff}^{2}$ (Fig. \ref{Invexr=0})
shows,in fact, that the correct function converges to a finite value.

\newpage

We now study the MA corrections to the radial motion of a particle in the Kerr metric. In
polar coordinates, the line element of this space--time can be written in the form given
by Boyer and Lindquist
 $$
 d\tau^2=\left( 1-\frac{2Mr}{\zeta^2} \right) dt^2-
 \frac{\zeta^2}{\Delta}dr^2-\zeta^2 d\theta^2 +
 $$
 \begin{equation}
 -\left(
 r^2+a^2+\frac{2Mr a^2}{\zeta^2}\sin^2 \theta \right)\sin^2 \theta
 d\varphi^2 +\frac{4M r a}{\zeta^2}\sin^2 \theta dtd\varphi\,,
 \label{Kerr line element}
 \end{equation}
where $\zeta^2=r^2+a^2 \cos^2 \theta$ and $\Delta=r^2-2Mr+a^2$. When $a$ vanishes, the
Schwarzschild geometry is recovered.

Three cases must be distinguished according to the possible values of $a$ and $M$. When
$a<M$, there are two null spherical surfaces of radii
\begin{equation}
r_{\pm}=M\pm\sqrt{M^2-a^2} \label{Horizons}\,.
\end{equation}
The external one is the event horizon. Particles can only enter, but not leave the
interior of this shell. In the region $r_{+}<r<r_{-}$, particles can only approach the
origin, while at radii smaller than $r_{-}$, particles are again allowed to move away
from the centre even if they can not re--emerge from $r_{-}$. Another characteristic
feature of this metric is that the event horizon does not coincide with the static limit.
This surface is given by the equation
\begin{equation}
r=M+\sqrt{M^2-a^2 \cos^2 \theta} \label{Static limit}
\end{equation}
and touches the event horizon at the two poles only. In the ergosphere the particles are
compelled to rotate around the centre since no timelike geodesics exist for constant
$\varphi$. For $a=M$ the two null surfaces $r_{+}$ and $r_{-}$ merge into a single
surface at $r=M$. The interior region is still inaccessible to external observers.
Finally, for $a>M$ there are no horizons and particles approaching the centre can always
return. However, the structure of the Kerr metric allows closed timelike geodesics that
violate causality for $a\geq M$ and inside $r_{-}$ for $a<M$ \cite{Carter,HawEll}.

We investigate the effects of MA in all three cases and restrict the motion, for
simplicity, to the  plane $\theta=\pi/2$. Use can be made of the integrals of motion
\cite{Carter,Landau}
\begin{eqnarray}
\dot{t} & = &-\frac{2M a}{r
\Delta}\tilde{L}+\frac{\tilde{E}}{\Delta}\left( r^2+a^2+\frac{2M
a^2 }{r} \right)\\%
\dot{\varphi} & = & \frac{\tilde{L}}{\Delta} \left( 1-\frac{2M}{r}
\right)+ \frac{2M a}{r \Delta} \tilde{E}\\%
\dot{r}^2 & = & \frac{1}{r^4} \left[ \left( r^2+a^2 \right) \tilde{E}-a \tilde{L}
\right]^2-\frac{\Delta}{r^4} \left[ \left(a \tilde{E}-\tilde{L} \right)^2+r^2 \right],
\end{eqnarray}
where $\tilde{E}$ and $\tilde{L}$ are the energy and angular momentum per unit of
particle mass . These expressions depend on $r$ only. In order to calculate the
components of $\ddot x^\mu$, it is sufficient to take their derivatives with respect to
$r$ and multiply them by $\dot{r}$. $\sigma^2$ can be constructed according to
(\ref{omegafirst}) and then used to determine the dynamics of a particle with MA
corrections. As explained above, only the first embedding needs to be considered.

The effective potential is defined by the equation
\begin{equation}
\left( \frac{dr}{ds}\right) ^{2}=\tilde{E}^{2}-V_{eff}^{2}\,,
\end{equation}
where the expressions for the momenta are derived, as usual
\cite{wh}, from the equation
\begin{equation}\label{gpp}
  \tilde{g}_{\mu\nu}p^{\mu}p^{\nu}=m^2\,,
\end{equation}
and the definitions $p^0=mdt/ds$, $p^1=mdr/ds$, $p^3=md\phi/ds$.
In (\ref{gpp}) $\tilde{g}_{\mu\nu}=\sigma^2g_{\mu\nu}$.

The expression of $V_{eff}^2$ for a particle moving in the
equatorial plane of Kerr space--time is
\begin{equation}
V_{eff}^{2}=\tilde{E}^{2}-\frac{\tilde{E}^{2}}{\sigma ^{4}\left(
r\right) } +\frac{1}{\sigma ^{2}\left( r\right) }\left(
1-\frac{2M}{r}\right) +\frac{\tilde{L}^{2}-a^2
\tilde{E}^2}{r^{2}\sigma ^{4}\left( r\right) } -\frac{2M
\left(\tilde{L}-a \tilde{E} \right)^2}{r^3 \sigma ^{4}\left(
r\right)} +\frac{a^2}{r^2 \sigma^2 \left(r \right)}\,.
\label{Effective potentialK}
\end{equation}
When MA tends to infinity, $\sigma^2(r) \to 1$ and the classical potential is recovered.
In the limit of vanishing angular momentum ($a\to 0$), one re-obtains the effective
Schwarzschild potential previously studied \cite{sch}.

\begin{figure}
\begin{center}
\resizebox{7cm}{!}{\includegraphics{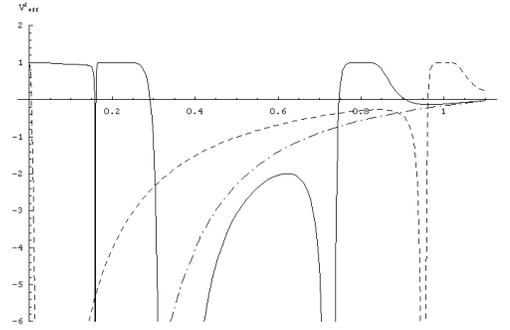}} \hfill
\caption{ \footnotesize {Potentials for $\tilde{E}=1$ and
$\tilde{L}=0$ in a Kerr space with $2M=1$ and $a=0.4$. Solid line:
Effective potential with MA corrections. Dashed line:
Schwarzschild potential with MA corrections. Dot-dashed line:
Classical potential.}} \label{A=0.4L=0}
\end{center}
\end{figure}

The results for each one of the cases listed above are the following.

a) $a<M$. Fig. \ref{A=0.4L=0}(drawn for the values $2M=1$, $a=0.4$, $\tilde{E}=1$,
$\tilde{L}=0$) shows that $\tilde{V}^2_{eff}$ is not modified at the static limit $r=1$,
while potential barriers form at $r_{+}=0.8$ and $r_{-}=0.2$. These barriers are a
consequence of the divergences of $\sigma^2$ at the horizons. One finds that
 \[
 \tilde{V}^2_{eff}(r)\sim \tilde{E}^2+\frac{4r_+^2(r_+-M)^2m^2}
 {M^2[a\tilde{L}+\tilde{E}(a^2+r_+^2)]^4}(r-r_+)^4
 \]
near $r_+$ and tends to $\tilde{E}^2$ at $r_+$. A particle coming from infinity cannot
therefore pass through the event horizon.

In the region $r_{-}<r<r_{+}$ each barrier is accompanied by a divergence that
corresponds to a zero in $\sigma^2$. One more divergence can be found near the origin
where the potential again approaches $\tilde{E}^2$.

\begin{figure}
\begin{center}
\resizebox{7cm}{!}{\includegraphics{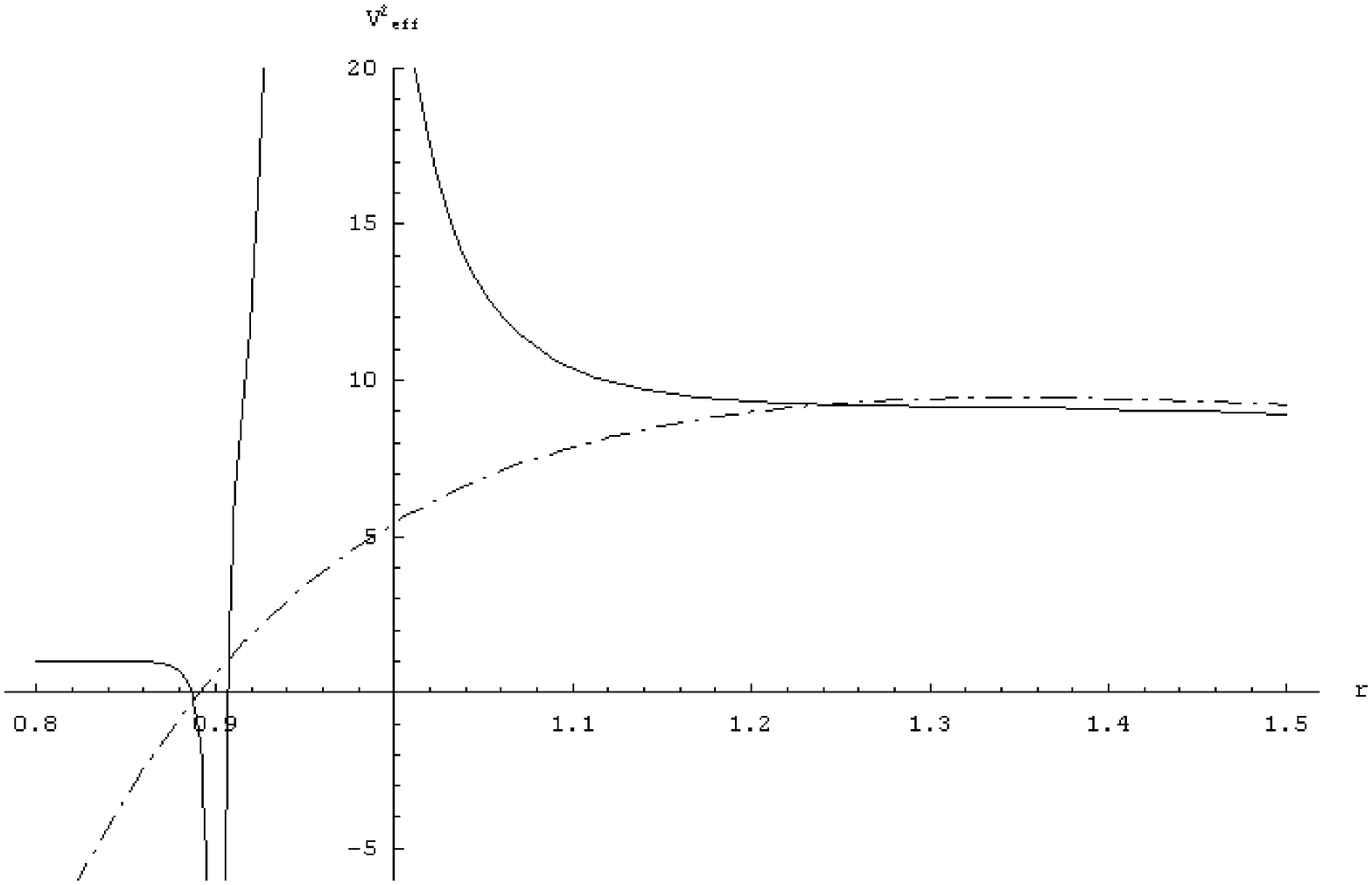}} \hfill
\caption{ \footnotesize {Potentials for $\tilde{E}=1$ and
$\tilde{L}=7$ in a Kerr space with $2M=1$ and $a=0.4$. Solid line:
Effective potential with MA corrections. Dot-dashed line:
Classical potential.}} \label{A=0.4L=7}
\end{center}
\end{figure}

\begin{figure}
\begin{center}
\resizebox{7cm}{!}{\includegraphics{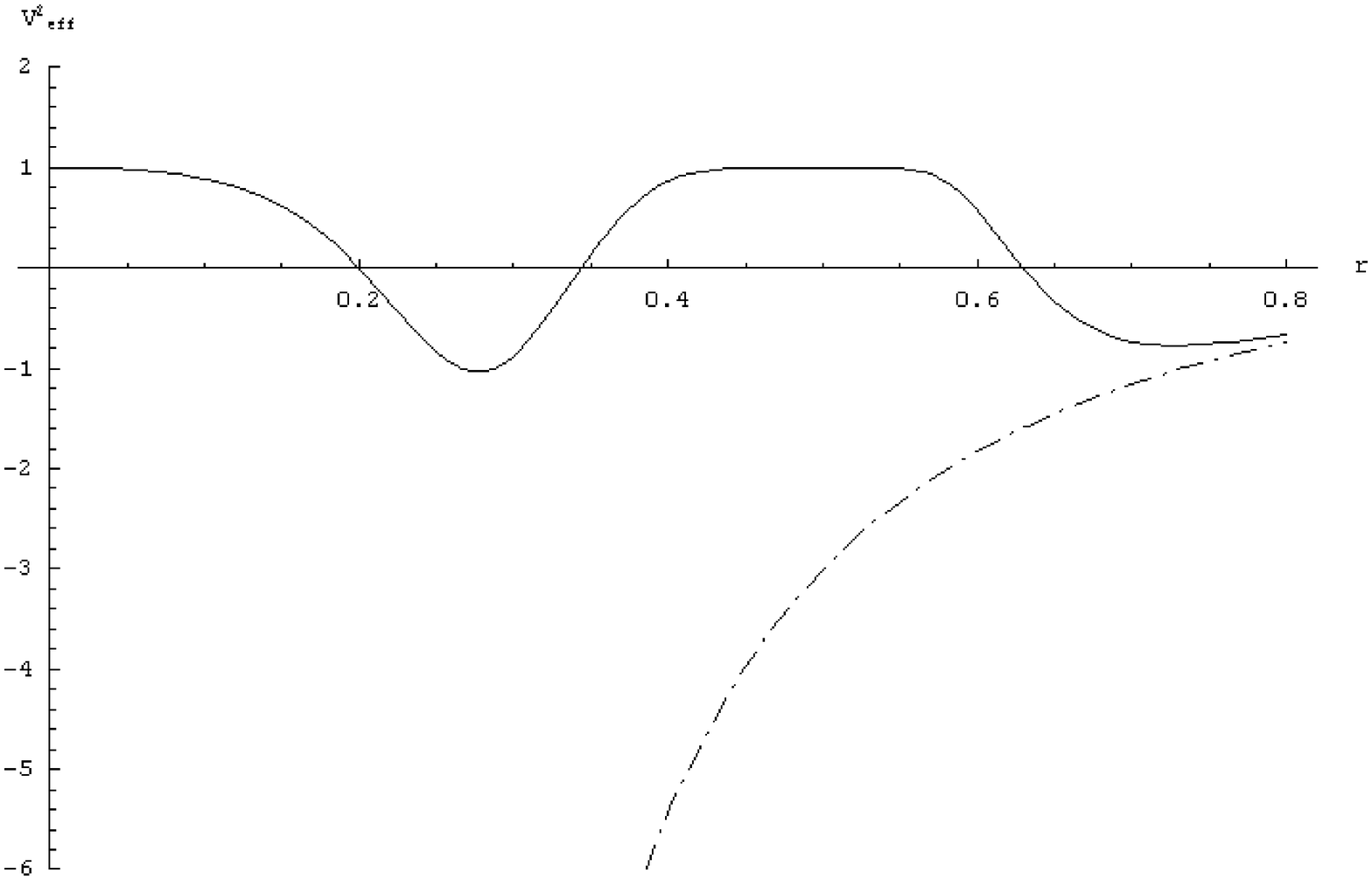}} \hfill
\caption{ \footnotesize {Potentials for $\tilde{E}=1$ and
$\tilde{L}=0$ in a Kerr space with $2M=1$ and $a=0.5$. Solid line:
Effective potential with MA corrections. Dot-dashed line:
Classical potential.}} \label{A=0.5L=0}
\end{center}
\end{figure}

Negative or low positive values of $\tilde{L}$ do not alter the
shape of the potential substantially. If $\tilde{L}$ is higher
than a certain threshold, two additional divergences appear
outside the event horizon (Fig. \ref{A=0.4L=7}). When
$\tilde{L}=a$, the classical potential has a positive (instead of
negative) divergence at the origin. $\tilde{V}^2_{eff}$ is always
regular at the origin, but the divergence near the origin becomes
positive on the right side.

b) When $a$ approaches M, the two horizons approach each other and so do the two barriers
generated by MA. The two divergences accompanying them merge and disappear and only the
two barriers are left, until they too merge when $a=M$ (Fig.\ref{A=0.5L=0}). In this case
all divergencies disappear and the potential is fully regularized.

\begin{figure}
\begin{center}
\resizebox{7cm}{!}{\includegraphics{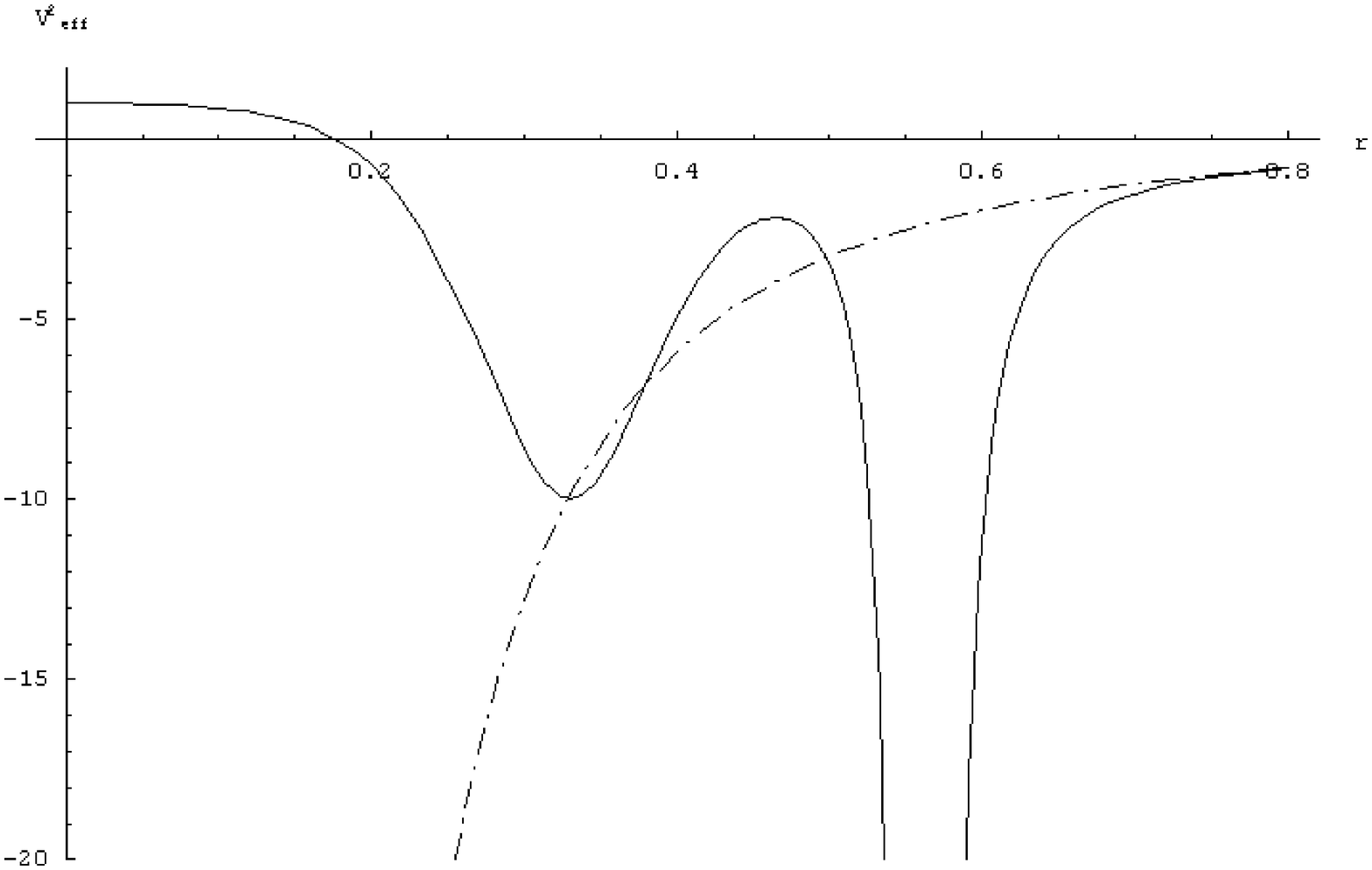}} \hfill
\caption{ \footnotesize {Potentials for $\tilde{E}=1$ and
$\tilde{L}=0$ in a Kerr space with $2M=1$ and $a=0.53$. Solid
line: Effective potential with MA corrections. Dot-dashed line:
Classical potential.}} \label{A=0.53L=0}
\end{center}
\end{figure}

We again obtain two divergences outside the horizon for high values of the angular
momentum. For $\tilde{L}=a$, we find a divergence near the origin as in a).

c) When $a > M$, there are no horizons and the barrier shrinks until it disappears. In
its place a negative divergence forms (Fig.\ref{A=0.53L=0}). The shape of the diagram
remains substantially unaltered even for $\tilde{L}\neq 0$. For the particular value
$\tilde{L}=a$, the effective potential is as in case b) (with $\tilde{L}=a$).

It is remarkable that MA has no effect on particles passing through the static limit. The
fact that particles in the ergosphere are bound to rotate does not lead to divergences in
the conformal factor. Then particles enter and leave this region as they normally would
in the absence of MA corrections.

Similarities in the structure of the Kerr and Reissner--Nordstr\"om \cite{reiss}metrics
are reflected in those of their effective potentials. There is a barrier on each null
surface with a divergence in the region between the two horizons. If the incoming
particle has an orbital angular momentum, nothing changes unless $L$ coincides with the
black hole angular momentum. Then the classical potential is strongly modified and the
effective potential changes, but the barriers at the event horizons remain. When $a > M$,
there are no more horizons and the barrier disappears, leaving a negative divergence that
would be accessible to the external particles.

These results have a bearing on what discussed in \cite{sch}. In fact, in physical
situations, collapsed bodies will likely have some angular momentum. One may then wonder
whether angular momentum perturbations invalidate the results obtained for the
Schwarzschild metric. This is not the case. The foregoing indicates that MA still
produces a barrier at the horizon, even though the rotation of the black hole modifies
the effective potential and the dynamics of the particles falling towards the event
horizon. The barrier tends continuously to that of the Schwarzschild  case and the fall
of particles is halted. Hence the black hole cannot absorb new matter. In the model, the
gravitational collapse of massive astrophysical objects is stopped before the occurrence
of the event horizon. A black hole does not therefore form, at least in the traditional
sense. However, a very compact radiating object would develop in its place, in appearance
very similar to a black hole.

A black hole would nonetheless form if the accreting matter were
first transformed into massless particles and these were absorbed
by the collapsing object at a rate higher than the corresponding
re-emission rate.

The results obtained represent a striking confirmation of and a decisive improvement upon
the conclusions reached in \cite{sch}. The iteration of the embedding approach does not
lead to a better approximation to the exact solution, because of the peculiar behaviour
of the equation relating a new conformal factor to the previous one. Yet this instability
does not affect the correctness of the first embedding which represents indeed the best
approximation to the exact solution.

This not only applies to small accelerations, but even reproduces qualitatively the
correct behaviour of the particle motion at the Schwarzschild radius. The occurrence of a
potential barrier at the gravitational radius is confirmed by the exact solution. The
analysis of the motion of a particle moving radially towards the origin indicates that
the proper time taken by the particle to reach the horizon is infinite. The particle
would never fall into the black hole.
The singularity in the approximate potential caused by a change of sign in $%
\sigma _{\left( 1\right) }^{2}$ is not present in the exact solution which is regular
even at the origin. Apart from the classical shift from 2M to $r_+$, the presence of
angular momentum in a black hole essentially leaves the barrier at the external horizon
unchanged. This means that all the remarks about the formation of black holes made in
\cite{sch} with regard to the Schwarzschild metric can be extended to that of Kerr. The
presence of the barrier would classically forbid, or at least slow-down, the formation of
a black hole.

Beside confirming the dynamics of collapsing objects, the application of Caianiello's
effective theory to the Kerr metric offers other interesting aspects. The structure of
$\tilde{V}^2_{eff}$ at the internal horizon $r_{-}$ is the specular image of that at
$r_{+}$. The intermediate region remains inaccessible from both sides. A barrier at a
horizon is always accompanied by a singularity on the side where $g_{00}$\ is negative.
These divergences are even present in the curvature invariants, so they must be
considered as physical singularities of the effective metric. Unlike the Schwarzschild
case, the singularity near the origin is always present because the repulsive effect of
the central object's angular momentum preserves the behaviour of $\sigma^2$ and
$\tilde{V}^2_{eff}$. For high values of the angular momentum the divergence in the
effective potential becomes positive and causes the formation of an infinite potential
barrier. Though physical, these singularities remain inaccessible to massive particles
and only regard the behaviour of matter inside the black hole, if the latter somehow
formed. In the case $a>M$  the barrier is directly accessible to particles coming from
infinite distances and could play an important role in their dynamics.

\bigskip
\bigskip
\begin{centerline}
{\bf Acknowledgments}
\end{centerline}
Research supported by MURST PRIN 99 and by the Natural Sciences
and Engineering Research Council of Canada.

\end{document}